\documentclass[pra,aps,twocolumn,nopacs,superscriptaddress,nofootinbib]{revtex4-1}

\usepackage{amsmath}  \usepackage{amssymb}  \usepackage{amsfonts}  \usepackage{bm}  \usepackage{bbm}   \usepackage{braket}   \usepackage{comment}  \usepackage{dcolumn}  \usepackage{enumerate}  \usepackage{epsfig}  \usepackage{gensymb}  \usepackage{graphicx}  \usepackage{indentfirst}  \usepackage{mathtools}  \usepackage{psfrag}  \usepackage{pst-all}  \usepackage{soul}  \usepackage{siunitx}   \usepackage{xcolor}  \usepackage{microtype} \usepackage{txfonts} \usepackage{txfontsb} 
\usepackage{float} 
\usepackage[colorlinks,linkcolor=blue,citecolor=blue,urlcolor=blue]{hyperref} 
\usepackage[T1]{fontenc} 

\def\ii{{\rm i}}  \def\ee{{\rm e}}
\def\me{m_{\rm e}}  
  
        \def\Eb{{\bf E}}                                      \def\rb{{\bf r}}      \def\vb{{\bf v}} 
    \def\zz{\hat{\bf z}}            
\def\kpar{k_\parallel}   
         
\def\Hh{\hat{\mathcal{H}}}  
  
\def\Sint{\hat{\mathcal{S}}^{\rm int}} 
\def\ah{\hat{a}}  
\def\ahd{\ah^{\dagger}}  

\begin{document}	

\title{Quantum sensing and metrology with free electrons}

\author{Cruz~I.~Velasco}
\affiliation{ICFO-Institut de Ciencies Fotoniques, The Barcelona Institute of Science and Technology, 08860 Castelldefels (Barcelona), Spain}
\author{F.~Javier~Garc\'{\i}a~de~Abajo}
\email{javier.garciadeabajo@nanophotonics.es}
\affiliation{ICFO-Institut de Ciencies Fotoniques, The Barcelona Institute of Science and Technology, 08860 Castelldefels (Barcelona), Spain}
\affiliation{ICREA-Instituci\'o Catalana de Recerca i Estudis Avan\c{c}ats, Passeig Llu\'{\i}s Companys 23, 08010 Barcelona, Spain}

\begin{abstract}
The quantum properties of matter and radiation can be leveraged to surpass classical limits of sensing and detection. Quantum optics does so by creating and measuring nonclassical light. However, better performance requires higher photon-number states, which are challenging to generate and detect. Here, we combine photons and free electrons to solve the problem of generating and detecting high-number states well beyond those reachable with light alone and further show that an unprecedented level of sensitivity and resolution is gained based on the measurement of free-electron currents after suitably designed electron--light interaction events. Our enabling ingredient is the strong electron--light coupling produced by aloof electron reflection on an optical waveguide, leading to the emission or absorption of a high number of guided photons by every single electron. We theoretically demonstrate that, by combining electron-beam splitters with two electron--waveguide interactions, the sensitivity to detect optical-phase changes can be enhanced tenfold using currently attainable technology. We further show that waveguided NOON states comprising tens of photons can be generated at megahertz rates based on electron post-selection after electron--waveguide interaction. These results inaugurate a disruptive quantum technology relying on free electrons and their strong interaction with waveguided light.
\end{abstract}
\date{\today}

\maketitle
\date{\today}

\section{Introduction}

Quantum sensing and metrology encompass a wide range of techniques that exploit the quantum properties of microscopic systems, such as entanglement and wave-like behavior, to achieve improved sensitivity and resolution compared to classical methods \cite{GLM06}. Quantum optics provides a powerful platform for implementing these ideas, where the generation and detection of nonclassical states of light have enabled measurements with unmatched sensitivity. For example, squeezed states played a key role in detecting gravitational waves, where a light interferometer was used to sense spatial modulations 18 orders of magnitude smaller than the wavelength of the lasers used \cite{A16}. More exotic states of light are also helpful in this context, such as the so-called NOON states, $[\ket{N,0}+\ket{0,N}]/\sqrt{2}$, where the first and second entries in each ket denote the numbers of quanta in two different channels (e.g., photons in different waveguides). NOON states represent maximally entangled superpositions of $N$-photon states that saturate the Heisenberg uncertainty limit \cite{D08}, enabling measurements with both super-sensitivity and super-resolution by beating the shot-noise and diffraction limits, respectively. At high $N$, these properties make NOON states powerful tools in quantum sensing and metrology \cite{D08,IRS14}, quantum lithography \cite{BKA00,DCS01}, and even quantum computing \cite{BL16}. However, high-$N$ NOON states (commonly referred to as high-NOON states) remain extremely challenging (e.g., $N=5$ generation is currently only possible at subhertz rates \cite{AAS10}).

In a separate effort, free electrons have recently emerged as powerful carriers of quantum information, which is encoded in their energy and propagation direction via states that can be prepared through interaction with both classical \cite{paper371} and quantum \cite{paper339} light using ultrafast electron microscopy techniques \cite{BFZ09,paper151,FES15,PRY17,B17_2,MB18_2,SMY19,RTN20,DNS20}. In addition, free electrons exhibit wave-like behavior that manifests in diffraction, interference, and superposition. These attributes have been exploited for the electron-based generation of quantum light. For example, following interaction with an optical cavity and electron post-selection, energy loss events experienced by individual electrons can herald the generation of quantum states of light, such as single \cite{paper180,FHA22} or multiple \cite{HEK23,AHF24} photon-number states. In addition, protocols have been proposed for the electron-based generation of squeezed \cite{paper402}, Gottesman--Kitaev--Preskill \cite{DBG23}, and entangled-pair \cite{paper430} photon states. Quantum entanglement between electrons and optical excitations has recently been demonstrated in two separate experiments \cite{HJS25,PBB25}, while quantum tomography has been applied to free electrons \cite{PRY17}, photoemitted electrons \cite{LLW25}, and optical cavity modes probed by free electrons \cite{GMK24}.

These emerging electron-based quantum technologies rely on the ability to couple electrons and light efficiently. However, the electron--photon interaction is generally low, and, for example, the number of photons created by a single electron in a single mode has remained below unity in experiments. To address this problem, several schemes have been proposed to increase such interaction \cite{K19,XCL24,KRR24,paper433}, and in particular, phase-matched coupling of electrons moving parallel and outside a waveguide \cite{paper180} constitutes a promising direction that has been argued to lead to the generation of high photon-number states \cite{paper433}. The combination of electron state superposition and strong electron--photon coupling can be anticipated to facilitate the generation of high-NOON states as well as the implementation of quantum sensing and metrology protocols that may potentially exceed the capabilities of all-optical schemes.

Here, we introduce free electrons as quantum probes on par with photons, enabling advanced quantum sensing and metrology protocols that extend the field well beyond its present frontiers. Specifically, we propose a class of systems that integrate free-electron wave optics with the interaction between free electrons and waveguided photonic modes. A central component in this approach is the design of realistic electron--light couplers capable of generating multiple photons per electron within a single photonic mode without creating electron-state decoherence. By leveraging this strong electron--light interaction and preparing electrons in coherent superpositions of different paths, we theoretically demonstrate protocols that enable both super-sensitivity and super-resolution in the measurement of optical phases via the detection of free-electron currents alone. We further combine strong coupling, path superposition, and energy-resolved electron detection to design systems capable of generating high-NOON states at unprecedentedly high rates (e.g., $N>10$ at megahertz rates). In summary, free electrons offer a powerful platform for quantum sensing and metrology, enabling previously unattainable capabilities through the integration of existing technologies.

\section{Results and Discussion}

\begin{figure*}[th!]
\centering\includegraphics[width=1.0\textwidth]{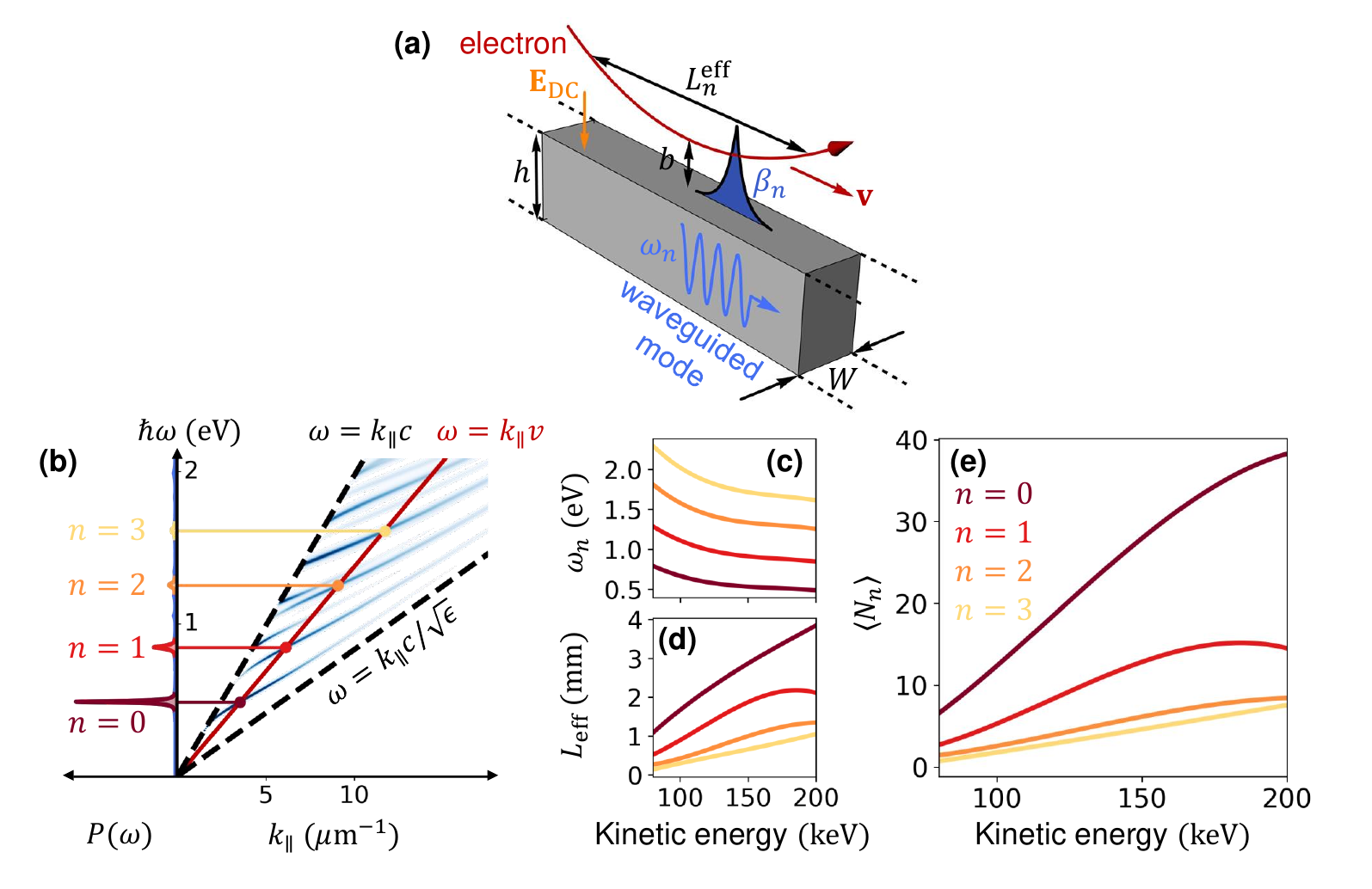}
\caption{\textbf{Multiple photon generation by individual free electrons.}
\textbf{(a)}~Sketch of an electron reflected at a grazing angle from an optical dielectric waveguide (rectangular cross section, width $W$, height $h$), following a bent trajectory with parallel velocity $\vb$, a minimum distance $b$ from the waveguide, and a mode-dependent effective interaction length $L^{\rm eff}_n$. The electron is repelled from the surface by a normal electric DC field $\Eb_{\rm DC}$. Waveguide modes with frequencies $\omega_n$ are excited with amplitudes $\beta_n\propto L^{\rm eff}_n$, provided they fulfill the phase-matching condition discussed in (b).
\textbf{(b)}~Mode dispersion in a diamond waveguide (permittivity $\epsilon=5.8$, $W=600$~nm, $h=800$~nm) as a function of photon wave vector $\kpar$ and frequency $\omega$. We plot the electron energy-loss probability (blue), revealing modes confined between the vacuum and bulk-diamond light lines (dashed lines) for photon energies below the material band gap (5.5~eV). The phase-matching condition $\omega=k_\parallel v$ is indicated in red for a $200$~keV electron ($v\approx0.7\,c$), together with the loss probability $P(\omega)$ (left curves, arbitrary units).
\textbf{(c-e)}~Frequency $\omega_n$ (c), effective interaction length $L^{\rm eff}_n$ (d), and average number of generated photons $\braket{N_n}$ (e) for the four dominant photonic modes [$n=1-4$, color-matched with the probability peaks in (b); see legend in (e)], plotted as functions of electron kinetic energy for $b=60$~nm and $E_{\rm DC}=10$~V/mm.
}
\label{Fig1}
\end{figure*}

\subsection{Efficient electron--photon coupler}

The key element enabling the present proposal for free-electron-based quantum sensing and metrology is the efficient generation and detection of quantum states of light through electron--photon interaction. To achieve this, we consider the configuration sketched in Fig.~\ref{Fig1}a, where an energetic free electron is incident at a grazing angle and reflects from a one-dimensional waveguide with rectangular cross section. We note that other waveguide geometries may also be suitable, provided the guide-mode fields extend beyond the material boundary, allowing the electron to couple to them. A repulsive DC electric field, oriented normal to the waveguide surface, is applied to prevent close collisions between the electron and the material. The minimum electron--surface distance $b$, determined by the incident glancing angle ($\ll1$), the electron kinetic energy, and the electric field amplitude $E_{\rm DC}$, is chosen such that inelastic excitations above the band gap of the material (e.g., electron-hole pairs) are neglibible, while coupling to guided modes remains sizeable. This is feasible because the fields of waveguided modes penetrate further into the vacuum, whereas those associated with high-energy inelastic excitations decay more rapidly with distance to the surface.

For simplicity, we consider a parabolic trajectory produced by a uniform repulsive field established between the waveguide (assumed to be slightly doped to act as a gate) and a nearby gate in the region near the waveguide. Given the small electron velocity along the perpendicular direction, we analyze the electron--photon coupling by considering parallel trajectories, in which the electron experiences an energy-loss probability $d\Gamma(x,\omega)/dz$. This quantity depends on the electron--surface distance $x$ and is normalized per unit length along the waveguide direction $z$. The probability is resolved in energy loss $\hbar\omega$, and we normalize it such that $\int_0^\infty d\omega\,[d\Gamma(x,\omega)/dz]=dP(x)/dz$ gives the total inelastic scattering probability per unit electron path length. We compute $d\Gamma(x,\omega)/dz$ from the self-induced field acting back on the electron \cite{paper149}. For rectangular waveguides, we use the boundary-element method \cite{paper040} (BEM) as an efficient numerical approach.

The electron maintains a constant velocity $\vb$ parallel to the waveguide, imposing a phase-matching condition $\omega=\kpar v$ between the frequency $\omega$ and the parallel wave vector $\kpar$ of the excitations that can be created in the waveguide \cite{paper180}. By varying the electron velocity to scan a wide $(\kpar,\omega)$ range, we reconstruct the dispersion diagram shown in Fig.~\ref{Fig1}b for a diamond waveguide. Here,  waveguide modes (blue features) appear within the region flanked by the vacuum and bulk-diamond light lines (dashed). For illustration, the phase-matching condition is indicated for 200~keV electrons (red line), along with the corresponding loss probability (left curves). Guided modes are well-defined at energies within the band gap of the material, where they exhibit zero linewidth. For visualization, we introduce a small imaginary component ${\rm Im}\{\epsilon\}$ in the permittivity, which broadens the modes into finite-width features. We define the excitation probability per unit length as $\int_n d\omega\,[d\Gamma(x,\omega)/dz]=dP_n(x)/dz$ for each waveguide mode $n$, where the frequency integral spans the corresponding peak in the energy-loss spectrum. This integral is approximately independent of ${\rm Im}\{\epsilon\}$ as long as this quantity remains small compared with $|\epsilon|$. Finally, we note that the deviation from the perfectly parallel trajectory in the glancing geometry leads to additional mode broadening, which we neglect in the following analysis.

The crossing points between the electron line $\omega=\kpar v$ and the guided modes determine the frequencies at which they are excited as a function of electron energy (Fig.~\ref{Fig1}c). Upon numerical inspection, the excitation probability per unit path length exhibits an exponential decay $dP_n(x)/dz\propto\ee^{-2x/\lambda_{\perp n}}$ as the electron moves away from the waveguide, consistent with the evanescent nature of the mode field outside the material. Based on the total light wave vector in vacuum ($\omega/c$), we anticipate the decay length to be approximately $\lambda_{\perp n}\approx(\kpar^2-\omega^2/c^2)^{-1/2}$, which increases for modes closer to the vacuum light line. This result is corroborated by fitting the BEM-calculated electron energy-loss probability as a function of distance $x$. The total excitation probability for each mode $n$ is obtained by integrating $dP_n(x)/dz$ over the parabolic electron trajectory $x=x_e(z)$, yielding $P_n=\int dz\,dP_n[x_e(z)]/dz$. From this, we define a mode-dependent effective interaction length $L^{\rm eff}_n$, such that $P_n=L^{\rm eff}_n[dP_n(x=0)/dz]$. After some algebra, we can express $L^{\rm eff}_n =\lambda_{\perp n}\,(v/c)\,\theta\,K_1(\theta)\,\exp(\theta-2b/\lambda_{\perp n})$ in terms of $b$, $E_{\rm DC}$, and $v$, where $K_1(x)$ is a Bessel function, $\me$ and $e$ are the electron mass and charge, respectively, and $\theta=2\me c^2\gamma/(eE_{\rm DC}\lambda_{\perp n})$.

Considering $b=60$~nm, $E_{\rm DC}=10$~V/mm, and a diamond waveguide (permittivity of 5.8) with a width of 600~nm and and height of 800~nm, the effective interaction lengths lie in the range of the millimeter range (Fig.~\ref{Fig1}d). Lower-order modes exhibit larger $L^{\rm eff}_n$ values, as they lie closer to the light line in the dispersion diagram (see Fig.~\ref{Fig1}a). In addition, $L^{\rm eff}_n$ increases with electron energy, since phase-matching occurs increasingly closer to the light line.

Starting from initially unpopulated waveguide modes, an excitation probability $P_n$ must be interpreted as a Poissian distribution of number-state populations in mode $n$ \cite{LKB1970,paper339,paper360} with an average number excited photons given by $\braket{N_n}=P_n=|\beta_n|^2$ (see below), where $\beta_n$ is the coupling amplidude \cite{paper339}
\begin{align} \label{betan} 
\beta_n=\frac{e}{\hbar\omega_n}\int_{-\infty}^{\infty}dz\;\zz\cdot\Eb_n(z)\,\ee^{-\ii\omega_nz/v},
\end{align}
which is proportional to the spatial Fourier transform of the normalized electric field $\Eb_n$ associated with mode $n$ along the $\vb=v\zz$ direction. Using BEM and integrated over the electron trajectory as explained above, we find relatively high values of $\braket{N_n}$ for the parameters under consideration, reaching up to $\braket{N_0}\sim 40$ photons in the lowest-order waveguide mode for 200~keV electrons, where phase-matching occurs at a photon energy of 0.48~eV.

\subsection{Quantum description of the electron--waveguide interaction}

For a highly collimated electron beam with a small lateral extent compared to the spatial variation scale of the optical fields to which it is exposed, the electron wave function can be factorized as $\psi^e_\parallel\psi^e_\perp$ (i.e., a product of components parallel and perpendicular to the electron velocity vector $\vb$). The details of $\psi^e_\perp$ are not relevant to our analysis, and therefore, we disregard them, except for the consideration of different electron paths.

We assume that the waveguide and the electron form a closed, lossless system, with energy transferred between the longitudinal motion of the electron and the guided modes, such that inelastic losses (e.g., due to electron-hole pair generation) are negligible. It is convenient to define creation and annihilation operators $a_n^\dagger$ and $a_n$ for the optical waveguide modes indexed by $n$. The photonic subsystem is then described in terms of photon-number states $\ket{\{N_n\}}$, where $N_n$ denotes the photon number in mode $n$ of frequency $\omega_n$. We also assume that the incident electron is prepared with a well-defined kinetic energy $\mathcal{E}_0$ (small broadening compared with the photon energies). From the assumption of energy conservation and the absence of losses, the electron energy changes to $\mathcal{E}_0-\hbar\sum_nN_n\omega_n$ upon the creation of a photon-number state $\ket{\{N_n\}}$. The electron state is thus uniquely defined once we know the photonic state, and therefore, we can represent the combined electron--light system just by $\ket{\{N_n\}}$. 

Transitions between these states are described by the interaction Hamiltonian \cite{paper339}
$\Hh_{\rm int} = -\ii e \sum_n \omega_n^{-1} \vb\cdot \big[\Eb_n(\rb) \ah_n -\Eb_n^*(\rb) \ahd_n\big]$, where $\Eb_n(\rb)$ is the electric field distribution associated with mode $n$. Starting from a given electron--light state $\ket{\psi(t\to-\infty)}$ before the interaction and following a previous analysis \cite{paper339}, the post-interaction state in the interaction picture reads \cite{paper339}
\begin{align} \label{psiSpsi} 
\ket{\psi(t\to\infty)}=\bigg[\prod_{n = 0}^{\infty} \ee^{\ii\chi_n}\;\Sint_n(\beta_n)\bigg]\ket{\psi(t\to-\infty)},
\end{align}
where
\begin{align} \label{intop} 
\Sint_n(\beta_n) =\ee^{-\beta_n\ah_n+\beta_n^{*}\ahd_n}
\end{align}
is a displacement operator characterized by a mode-dependent coupling amplitude $\beta_n$ [Eq.~({\ref{betan})], while the phase $\chi_n$ can be absorbed by a global phase associated with each electron path. Under the conditions considered in this work, only a few low-order modes $n$ contribute efficiently to Eq.~(\ref{psiSpsi}). For a waveguide initially prepared in the photonic vacuum state, the final population of mode $n$ is described by
\begin{align} \label{entcohst} 
\Sint_n(\beta_n) \ket{0} &= \ee^{-|\beta_n|^2/2}\sum_{N_n=0}^\infty \frac{(\beta_n^*)^{N_n}}{\sqrt{N_n!}} \ket{N_n},
\end{align}
which prescribes a Poissonian distribution of $\ket{N_n}$ states with an average number of created photons given by $\langle N_n\rangle=|\beta_n|^2$.

\begin{figure}
\centering\includegraphics[width=.35\textwidth]{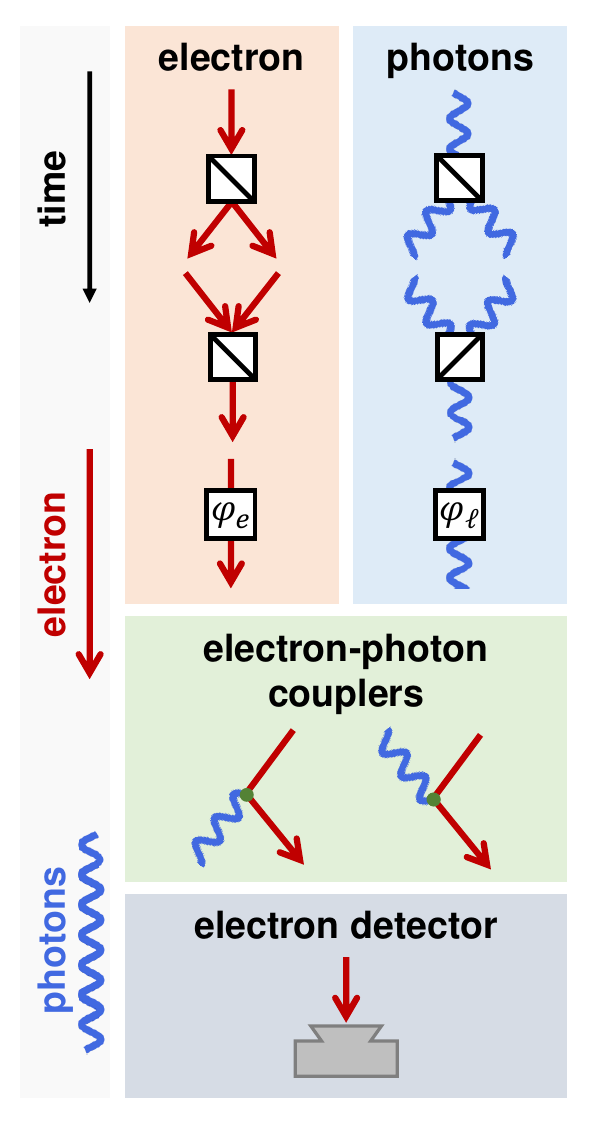}
\caption{\textbf{Building blocks of quantum free electronics.} QUAFE components include beam splitters/mixers and phase shifters for electrons (phase $\varphi_e$) and waveguided photons ($\varphi_\ell$), as well as electron--photon couplers (for light emission/absorption; left/right) and electron detectors. Time flows downward in each diagram. Electrons and photons are represented by red arrows and blue waves, respectively.}
\label{Fig2}
\end{figure}

\subsection{Quantum free electronics}

The strong interaction between free electrons and waveguided photons can be harnessed to generate NOON states and implement quantum sensing and metrology protocols within the framework of quantum free electronics (QUAFE). This approach involves integrating the electron--photon couplers discussed above with additional components for manipulating light and electrons, such as beam splitters, phase shifters, and electron detectors, as illustrated in Fig.~\ref{Fig2}. In this work, we combine these elements to develop practical quantum sensing and metrology schemes that achieve super-sensitivity and super-resolution, relying solely on electron current measurements.

While splitters, mixers, and phase shifters for photons are standard in optical setups, analogous elements exist for free electrons in the context of electron microscopy. For example, the electron biprism was developed seven decades ago \cite{MD1956} and can function as a beam splitter: a biased wire placed in the path of an electron plane wave deflects electrons on either side into two paths with well-defined angles \cite{STP17}. A more versatile alternative is provided by transmission gratings, which have been shown to split electron waves into different diffraction orders \cite{JTM21}. By selecting two dominant diffraction directions, they can be used to implement beam splitting and mixing, enabling the realization of a free-electron Mach-Zehnder interferometer \cite{JTM21} as well as the measurement of optical properties in a specimen \cite{paper388}. In what follows, we consider gratings that split an incident plane wave into two dominant diffraction directions with identical transmission amplitudes. Likewise, we consider gratings that mix two beams into a single one, with each incident beam contributing an identical amplitude. Efficient electron gratings have been demonstrated with amplitudes $\sim1/4$ or higher \cite{JTM21}, which introduce an overall correction factor in our results without involving a substantial decrease in electron current.

The electron phase shifter in Fig.~\ref{Fig2} operates by multiplying the electron wave function by a phase factor $\ee^{\ii\varphi_e}$, while a phase $\varphi_\ell$ applied to an optical waveguide mode $n$ introduces a factor $\ee^{\ii N_n\varphi_\ell}$ in the number state with $N_n$ photons. Since we are interested in sensing small optical phases, we assume a mode-dependent phase shift of the form $\varphi_\ell\omega_n/\omega_0$, proportional to the mode frequency (as would result from transmission through a thin dielectric film), with $\varphi_\ell$ representing the phase shift for the fundamental mode $n=0$.

The operations described in this section constitute the building blocks of QUAFE systems, which can be represented by diagrams that combine the elements shown in Fig.~\ref{Fig2}, arranged vertically to follow the downward evolution of time and concluding with electron detection (i.e., we focus on protocols based solely on electron measurements). When the electron reaches the detector, the system collapses into a specific electron energy state, or equivalently, a well-defined photonic state due to the aforementioned electron--photon entanglement, assuming monochromatized incident electrons (see above). This property is central to the protocols presented next.

\begin{figure*}
\centering\includegraphics[width=0.95\textwidth]{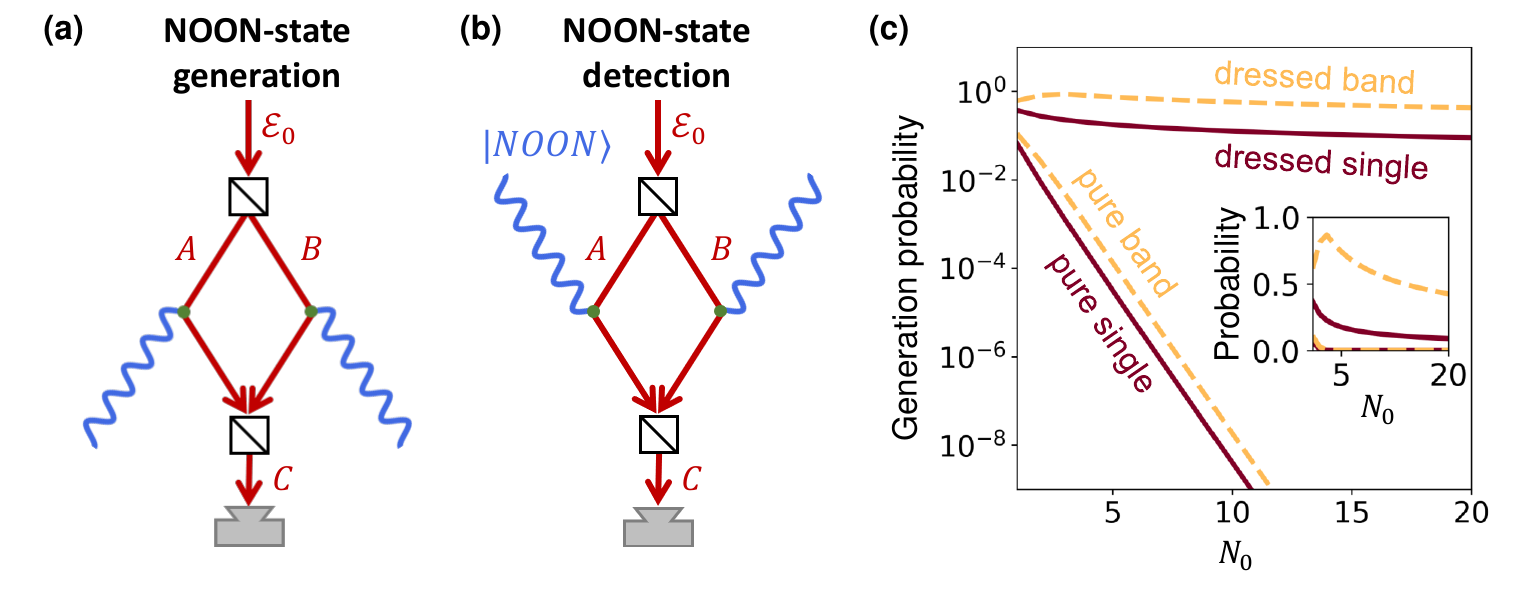}
\caption{\textbf{High-NOON state generation.}
\textbf{(a)}~QUAFE configuration for generating NOON states consisting of $N_0$ photons in the $n=0$ waveguide mode and no photons in modes with $n\neq0$, heralded by the detection of an electron that has lost energy $N_0\hbar\omega_0$ relative to the incident energy $\mathcal{E}_0$. Electron paths $A$ and $B$ interact with two separate waveguides and are subsequently combined into a single path $C$, where electron detection takes place.
\textbf{(b)}~QUAFE configuration for detecting NOON states. Using the same electron-path configuration as in (a), paths $A$ and $B$ interact with separate waveguided components of the NOON state. The detection probability depends on the interference between the two electron paths and the relative phase $\varphi_\ell$ between the two NOON-state components.
\textbf{(c)}~Probability of generating the waveguided NOON state discussed in (a) as a function of $N_0$, normalized to the number of transmitted electrons. We compare the probability of producing a pure $N_0$ state (pure single) with the cumulative probability of producing $N_0$, $N_0\pm1$, and $N_0\pm2$ states (pure band), under the condition that no other waveguide modes (i.e., $n\neq0$) are excited. We also show the corresponding probabilities when excitation in modes $n\neq0$ is allowed (dressed single and band). The inset shows the same plot on a linear vertical scale. The waveguide parameters are the same as in Fig.~\ref{Fig1}, with fixed $\mathcal{E}_0=200$~keV and $b=60$~nm, while $L_0^{\rm eff}$ is optimized to maximize the probability of the target $N_0$.}
\label{Fig3}
\end{figure*}

\subsection{Efficient generation of high-NOON states}

We first explore the application of QUAFE to generate them with a high photon number. This can be achieved using the configuration shown in Fig.~\ref{Fig3}a, where the incident electron is split into two paths $A$ and $B$, but now, each of them interacts with a different waveguide. These paths are subsequently recombined into a common path $C$, where electrons are detected with energy resolution much finer than the energy of the generated photons. As explained above for interactions with a single waveguide, the detection of an electron that has lost an energy $\hbar\sum_nN_n\omega_n$ heralds the creation of a photon-number state $\ket{\{N_n\}}$. In the present configuration, the electron mixer erases which-path information, resulting in the creation of a superposition state  $\ket{\{N_n\},\{0\}}+\ket{\{0\},\{N_n\}}$, where the two entries in each ket refer to the two different waveguides.

The probability of creating a pure NOON state in the waveguide modes $n=0$ (i.e., with $N_0\neq0$ and $N_n=0$ for all $n\neq0$) is given by the product $P_{0,N_0}P_{1,0}P_{1,0}\cdots$, where $P_{n,N}=\ee^{-P_n}P_n^N/N!$ is a Poissonian distribution of photon-number states created in mode $n$, and $P_n$ is the average photon population. This {\it pure} NOON-state generation probability decreases rapidly with $N_0$, as shown in the lowest curve of Fig.~\ref{Fig3}c. In this plot, the waveguide parameters are the same as in Fig.~\ref{Fig1}, but the effective interaction length is optimized to maximize $P_{0,N_0}$. In a more practical scenario, excitation of higher-frequency photons ($\omega_{n\neq0}>\omega_0$), which {\it dress} the photonic state of the system, can be tolerated, as they can be filtered out after generation. Then, the NOON-state generation probability is simply $P_{0,N_0}$, which reaches values between 0.1 and 1 for $N_0$ up to 20. For typical beam currents of $10^9$ electrons per second in electron microscopes, this probability translates into a NOON-state generation rate of $\sim10^8$~Hz for $N_0\sim10-20$.

As a complement to electron-based NOON-state generation, a similar scheme could be envisioned for NOON-state detection (Fig.~\ref{Fig3}b), where two waveguides supporting the NOON state are probed by separate electron paths. However, efficient electron--light interaction requires precise temporal overlap between the photons and the probing electron, demanding challenging temporal synchronization and spectral shaping of the electron wave packet. In addition, the probability for the electron absorbing a large number of photons decreases rapidly (analogously to that in pure NOON-state generation), rendering this approach experimentally impractical. Instead, photon detection and generation by an individual electron within a QUAFES setup is more advantageous, as we show next, eliminating the need for certified NOON-state generation and surpassing the performance of more traditional optics-based methods.

\begin{figure*}
\centering\includegraphics[width=0.95\textwidth]{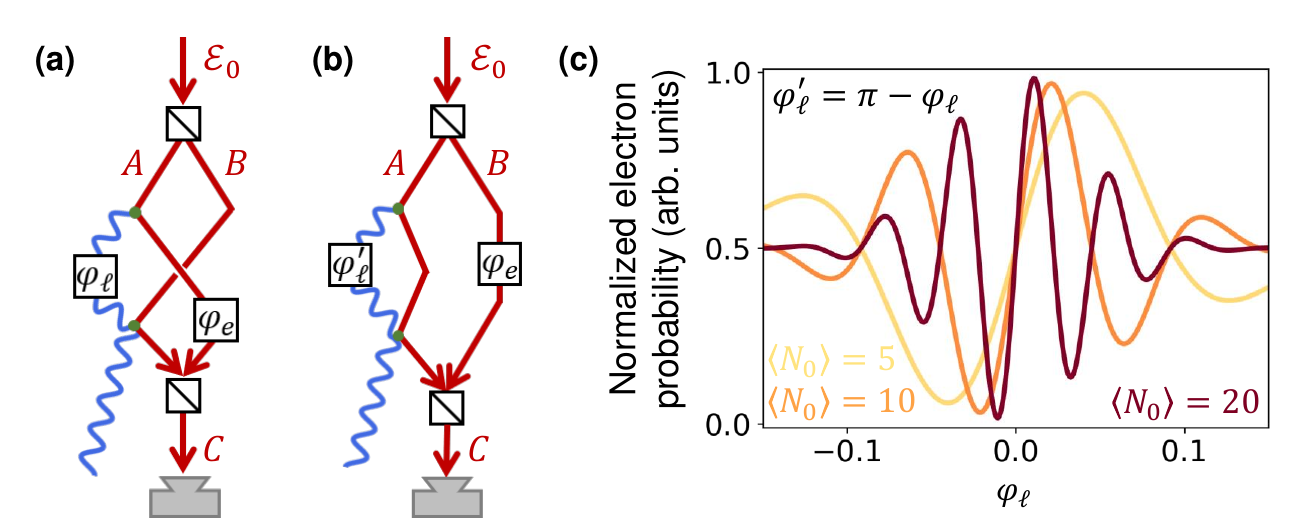}
\caption{\textbf{Super-sensitivity and super-resolution through QUAFE.}
\textbf{(a)}~Proposed configuration for super-sensitive phase measurements. An electron wave is split into two paths, $A$ and $B$, which are later mixed into a common path $C$. Paths $A$ and $B$ interact with a single optical waveguide. An optical phase $\varphi_\ell$ is introduced between the two electron--light interaction regions. The electron current is measured at the output of path $C$. A controllable electron phase is applied to path $B$ to adjust the relative phase difference between paths $A$ and $B$. Different elements of Fig.~\ref{Fig2} are combined in this protocol.
\textbf{(b)}~Alternative configuration in which only path $A$ interacts with the waveguide, undergoing two separate interactions in different regions.
\textbf{(c)}~Normalized electron detection probability for the configuration shown in (a) as a function of the optical phase $\varphi_\ell$ introduced in waveguide mode $n=0$. The phases of higher-order modes are assumed to scale proportionally with their respective frequencies. Different curves correspond to selected values of the average photon number in mode $n=0$ ($\braket{N_0} = |\beta_0|^2$), under the conditions of Fig.~\ref{Fig1}c-e for $200$~keV electrons. The same plot also applies to the configuration in (b) by replacing $\varphi_\ell$ with $\pi-\varphi_\ell'$.}
\label{Fig4}
\end{figure*}

\subsection{Super-sensitivity and super-resolution in QUAFE phase measurements}

We present two examples of protocols that enhance sensitivity and resolution in the measurement of an optical phase (Fig.~\ref{Fig4}a,b). In Fig.~\ref{Fig4}a, an incident electron is split into two paths ($A$ and $B$), each interacting once with the same waveguide. These two interactions are configured identically, so that the corresponding coupling coefficients $\beta_n$ are equal. The optical phase $\varphi_\ell$ to be measured is introduced in the waveguide modes between the two interaction regions. The electron paths are then recombined into a single path $C$, where the electron current is measured. In the absence of electron--photon coupling, the setup functions as a Mach-Zehnder interferometer, in which we also introduce a controllable phase shift $\varphi_e$ between the two electron paths. Analyzing the electron current $I$ using the formalism introduced above, we find
\begin{align} \label{int1} 
I\propto 1+&\exp\Big[-\sum_n \braket{N_n}\big(1-\cos\varphi_{\ell n}\big)\Big] \\
&\times\cos\bigg[\varphi_e+ \sum_n \braket{N_n} \sin\varphi_{\ell n}\bigg], \nonumber
\end{align}
where $\varphi_{\ell n}=\varphi_\ell\omega_n/\omega_0$ (see Methods). The electron current oscillates with $\varphi_\ell$ in a manner dependent on the average number of generated photons, as illustrated in Fig.~\ref{Fig4}c for $\varphi_e=\pi/2$. This current modulation arises from electron scattering at the mixer into undetected channels away from path $C$, depending on the phase difference between electron paths $A$ and $B$, which is governed by the interaction with the waveguide (see Methods). As anticipated, when $\braket{N_0}\gg1$, the current exhibits sharp oscillations, enabling super-resolution in the measurement of small phase variations. In addition, assuming $\varphi_\ell\ll1$, Eq.~(\ref{int1}) reduces to $I\propto1+\cos(\varphi_e+N^{\rm eff}\varphi_\ell)$, where $N^{\rm eff} = \sum_{n} \braket{N_n}\,\omega_n/\omega_0$ represents an effective photon number that takes into account the weighted contributions from all waveguide modes. For the parameters considered in Fig.~\ref{Fig4}c, we obtain $N^{\rm eff} \approx 7\times \braket{N_0}$, which explains the sharp features observed in the electron current.

The sensitivity of this method can be estimated from the maximum slope of the measured current, evaluated as $\partial I/\partial \varphi_\ell$ at $\varphi_\ell=0$, which reveals a linear dependence of current variations on both optical phase variations and the effective photon number: $\Delta I \propto N^{\rm eff}\Delta\varphi_\ell$.

In an alternative configuration, both electron--photon interactions can take place along the same path $A$ (Fig.~\ref{Fig4}b), leading to analogous results as in the previous configuration, so that Fig.~\ref{Fig4}c still applies, but with $\varphi_\ell$ substituted by $\pi-\varphi_\ell'$ in the horizontal axis.

Importantly, the proposed configurations require only electron current measurements (i.e., neither photon detection nor electron energy resolution is necessary). This makes the technique robust against variations in incident electron energy, provided the coherent energy spread of individual electrons is smaller than the photon energies involved. This condition is met by low-coherence electron sources, ensuring that different photon-number states created by the electron do not interfere once the electron is traced out.

\section{Conclusions}

In summary, we capitalize on the quantum nature of free electrons to theoretically demonstrate optical phase measurements with enhanced sensitivity and resolution, relying solely on the detection of electron currents. Central to this approach is an efficient electron--photon coupling scheme, which we achieve via grazing-angle reflection of energetic electrons from electrically biased one-dimensional waveguides. By coherently splitting the electron into a superposition of two paths, we show that this coupling scheme enables the efficient generation of NOON states with high photon numbers (e.g., $N=10-20$ at $10^8$~Hz rates). Although NOON-state detection remains a challenge, we bypass this problem by formulating an alternative metrology strategy in which individual electrons both generate and probe photonic states, thus eliminating the need for photon detection. We exploit the interference between photon emission and absorption processes by the electron to probe the optical phase, benefiting from the effect of phase amplification associated with large photon-number states. Importantly, these states are mutually incoherent because they are entangled with different final electron energies. We conclude that the integration of electron wave splitting, recombination, and controlled coupling to photons constitutes a practical and powerful platform for quantum free electronics, holding strong potential to revolutionize the field of quantum sensing and metrology through the unique quantum properties of free electrons.

\section*{Methods}

\vspace{8px} \noindent {\bf Derivation of equation (\ref{int1}).} We are interested in the transmitted electron current in the configuration of Fig.~\ref{Fig4}a. After each electron path $A$ and $B$ interacts with a waveguide initially prepared in the photonic vacuum state, the joint electron--light system evolves into the state described by Eq.~(\ref{entcohst}). Considering a single-mode waveguide for simplicity, the electron--light state can be written as $\ee^{-|\beta|^2/2}\sum_{N=0}^\infty (\beta^*)^{N}(N!)^{-1/2} \big[\ee^{\ii(\varphi_e+N\varphi_\ell)}\ket{A}+\ket{B}\big]\otimes\ket{N}$, where $\ket{A}$ and $\ket{B}$ represent the two electron paths, the photon-number state $\ket{N}$ is shared by both paths (i.e., there is only one waveguide), and we include the electron and optical phases $\varphi_e$ and $\varphi_\ell$ introduced in path $A$. Note that the optical phase $\varphi_\ell$ needs to be multiplied by the photon number $N$. As discussed above, the mode-dependent phase is taken as $\varphi_\ell\omega_n/\omega_0$. The two paths are now recombined into a single path $C$ (see Fig.~\ref{Fig4}a). Assuming equal transmission coefficients $A\to C$ and $B\to C$, and extending the model to include all waveguide modes $n$, the amplitude of the component associated with path $C$ and a photon-number state $\ket{\{N_n\}}$ becomes $\propto\big[\ee^{\ii\phi_{\{N_n\}}}+1]\prod_n\ee^{-|\beta_n|^2/2}(\beta_n^*)^{N_n}(N_n!)^{-1/2}$, where $\phi_{\{N_n\}}=\varphi_e+\varphi_\ell\sum_n N_n\omega_n/\omega_0$ is the overall phase accummulated by path $A$ (first term inside the square brackets) relative to path $B$ (second term). Finally, Eq.~(\ref{int1}) is readily obtained from the squared modulus of this amplitude after summing over all $\{N_n\}$ combinations.

\section*{ACKNOWLEDGEMENTS}

This work has been supported in part by the European Research Council (101141220-QUEFES), the Spanish MICINN (Severo Ochoa CEX2019-000910-S), the Catalan CERCA Program, and Fundaci\'os Cellex and Mir-Puig.


%

\end{document}